\documentclass[conference]{IEEEtran}
\IEEEoverridecommandlockouts
\usepackage{cite}
\usepackage{amsmath,amssymb,amsfonts}
\usepackage{algorithmic}
\usepackage{graphicx}
\usepackage{textcomp}
\usepackage{xcolor}
\def\BibTeX{{\rm B\kern-.05em{\sc i\kern-.025em b}\kern-.08em T\kern-.1667em\lower.7ex\hbox{E}\kern-.125emX}}
    
\usepackage{tikz}
\usetikzlibrary{dsp,chains}
\usetikzlibrary{arrows}
\usetikzlibrary{quotes,arrows.meta}

\DeclareSymbolFont{matha}{OML}{txmi}{m}{it}
\DeclareMathSymbol{\varv}{\mathord}{matha}{118}

\DeclareMathAlphabet{\mathcal}{OMS}{cmsy}{m}{n}
\DeclareMathAlphabet\mathbfcal{OMS}{cmsy}{b}{n}

\def\EDC{\mathrm{EDC}}

\def\EDD{\mathrm{EDD}}

\def\g{\mathbf{g}}

\def\DeltaPerSampleAtt{\g_{i}}

\usepackage{orcidlink}
\usepackage{hyperref} 

\newcommand\copyrighttext{%
  \footnotesize This work has been accepted for the I3DA 2021 International Conference and will be submitted to IEEE Xplore Digital Library for possible publication. Copyright may be transferred without notice, after which this version may no longer be accessible.
  }

\newcommand\copyrightnotice{%
\begin{tikzpicture}[remember picture,overlay]
\node[anchor=south,yshift=10pt] at (current page.south) {\fbox{\parbox{\dimexpr\textwidth-\fboxsep-\fboxrule\relax}{\copyrighttext}}};
\end{tikzpicture}%
}

\begin{document}

\copyrightnotice

\title{A Method for Capturing and Reproducing Directional Reverberation in Six Degrees of Freedom
\thanks{This work has been funded by the Kaute Foundation (project "Conservation of the Acoustics of Historical Halls in Helsinki"). This study is part of the activities of the ``Nordic Sound and Music Computing Network---NordicSMC'', NordForsk project number 86892. }}
 
\author{\IEEEauthorblockN{Benoit Alary\,\orcidlink{0000-0002-3825-4941}}
\IEEEauthorblockA{\textit{Acoustics Lab,} \\
\textit{Dept. of Signal Processing and Acoustics} \\
\textit{Aalto University}\\
Espoo, Finland}
\and
\IEEEauthorblockN{Vesa V\"{a}lim\"{a}ki\,\orcidlink{0000-0002-7869-292X} \IEEEmembership{Fellow, IEEE}}
\IEEEauthorblockA{\textit{Acoustics Lab,} \\
\textit{Dept. of Signal Processing and Acoustics} \\
\textit{Aalto University}\\
Espoo, Finland}}

\maketitle

\begin{abstract}
The reproduction of acoustics is an important aspect of the preservation of cultural heritage. A common approach is to capture an impulse response in a hall and auralize it by convolving an input signal with the measured reverberant response. For immersive applications, it is typical to acquire spatial impulse responses using a spherical microphone array to capture the reverberant sound field. While this allows a listener to freely rotate their head from the captured location during reproduction, delicate considerations must be made to allow a full six degrees of freedom auralization. Furthermore, the computational cost of convolution with a high-order Ambisonics impulse response remains prohibitively expensive for current real-time applications, where most of the resources are dedicated towards rendering graphics.
For this reason, simplifications are often made in the reproduction of reverberation, such as using a uniform decay around the listener. However, recent work has highlighted the importance of directional characteristics in the late reverberant sound field and more efficient reproduction methods have been developed.
In this article, we propose a framework that extracts directional decay properties from a set of captured spatial impulse responses to characterize a directional feedback delay network. For this purpose, a data set was acquired in the main auditorium of the Finnish National Opera and Ballet in Helsinki from multiple source-listener positions, in order to analyze the anisotropic characteristics of this auditorium and illustrate the proposed reproduction framework.
\end{abstract}


\section{Introduction}
Large historical buildings, such as churches, amphitheaters, and concert halls, all carry acoustic properties leading to invaluable reverberation, unique for each hall. For this reason, capturing the acoustics of these spaces plays an essential role in the conservation of their cultural heritage. Acquiring a detailed representation of their acoustics is also a key aspect in reproducing these spaces in mixed reality (MR) applications, such as virtual reality (VR) and augmented realty (AR), where a listener may move freely in an environment reproduced virtually. 

In order to auralize acoustics in real time, most reproduction methods are required to make a compromise between accuracy and efficiency. For instance, as the perceptual importance of salient early reflections (ERs) is well known \cite{Barron_ER_1971}, special care is usually given to their reproduction \cite{Schroeder_simulation_1970}, while the late reverberation is often reduced to decaying stochastic noise \cite{polack_playing_1993}. Indeed, while never fully realized in practice, the diffusion of sound energy in a room over time tends to lead to a more homogeneous and isotropic distribution \cite{nolan_isotropy_2020}. In an opera hall, the orchestral pit and the large area behind the main stage form a coupled volume with the seating area of the hall and their distinct decay characteristics creates an anisotropic sound field \cite{Massimo_opera_2016}.

One common method for the reproduction of acoustics is to capture room impulse responses (RIRs) or spatial room impulse responses (SRIRs) \cite{gerzon_recording_1975, farina_recording_2003}. While capturing SRIRs is the most accurate way to preserve the acoustics properties of a hall, this approach requires many measurement positions when a sound field is inhomogeneous \cite{Zotter_Aur_2020}.
Another approach to auralize the acoustics of a hall is to produce its response artificially using virtual acoustics methods \cite{Noisternig_framework_2008, Katz_Reconstruction_2020}.

Delay network reverberators are a different type of reproduction methods used to reproduce the perceived characteristics of reverberation \cite{jot_digital_1991, valimaki_fifty_2012}. In one of their earliest form, a simple multi-tap recirculating delay line was used to enhance the acoustics properties of a hall \cite{Vermeulen_stereo_1958}. More complex delay networks emerged once the cost of physical delay units decreased and with the introduction of the all-pass filter to produce a colorless reverberation effect \cite{schroeder_colorless_1961}.
Delay networks continued to evolve, and a more generalized design came with the feedback delay network (FDN) \cite{jot_digital_1991}, still commonly used today. 

In multichannel reproduction, delay networks are commonly used to create sets of decorrelated signals, each decaying towards a unique reverberation time, thus assuming an isotropic sound field. However, rooms lacking good diffusion properties exhibit inhomogeneous and anisotropic decay of sound \cite{Waterhouse_Interference_1955, nolan_isotropy_2020}. Recent work has proposed novel ways to analyze decaying sound fields \cite{alary_assessing_2019, nolan_wavenumber_2018, berzborn_directional_2019, Berzborn_dsfda_2021} while directional decay characteristics have also been found to be perceptually important once a certain threshold is met \cite{romblom_perceptual_2016, Alary_perceptual_2021}.

Recently, modified versions of the FDN were proposed to produce direction-dependent reverberation characteristics using a directional feedback delay network (DFDN) \cite{alary_directional_2019, alary_frequency-dependent_2020}.
In this article, we propose a method to extract key directional characteristics from a set of captured SRIRs to produce a compact data set of properties used to specify the parameters of a DFDN for auralization. The reverberation algorithm also offers six degrees of freedom (6DoF) auralization by interpolating between the values in the data set to modulate the gains in the delay network. 

In the next section, we review the analysis of directional decay characteristics and their reproduction using a DFDN. Section III describes the proposed reproduction framework and provides analysis results from the data set of SRIRs measured in the main auditorium of the Finnish National Opera and Ballet in Helsinki. Section IV concludes the paper and discuss future work.

\section{Background}

\subsection{Statistical Sound Field Analysis}

A captured RIR contains time and frequency information on the decay of energy in a room, but the recording is also susceptible to environmental noise present during the recording. Early in a RIR, sparse ERs are present. However, after a short period of time, the density of echoes in the RIR is such that it is no longer possible to identify individual reflections \cite{schroeder_colorless_1961}. For this reason, decay properties of a late reverberant sound field are better analyzed through statistical methods, such as the energy-decay curve (EDC) \cite{schroeder_new_1965}, which consists of the integration of energy from a given sample $n$ until the end of the measured RIR, as defined in dB by
\begin{equation}
\EDC(n) = 10\,\log_{10}\Bigg(\sum_{i=n}^{N} \big(y(i)\big)^2\Bigg),
\label{eq:DEDC}
\end{equation}

\noindent where $N$ is the length of the RIR in samples. Since the decay is rarely uniform at all frequencies, it is preferable to analyze the decay as a frequency-dependent phenomenon. For this purpose, the energy-decay relief (EDR) \cite{jot_analysissynthesis_1992} extends the EDC analysis by using a filter bank to obtain a set of frequency-dependant decay curves.

A common descriptor of late reverberation is the decay time, usually defined as the time required to reach a 60~dB energy decay ($T_{60}$). Due to the rapid drop of energy in the beginning of the response and the presence of noise  \cite{masse_denoising_2020}, a $T_{60}$ is usually estimated from a $T_{30}$, measured from a 30~dB decay following the first 10~dB drop, called the early decay time (EDT) \cite{Hopkins_Insulation_2007}. 

To analyze a spherical sound field, these measures of energy decay are expanded to a set of angle-dependent directional room impulse responses (DRIRs). More specifically, using a captured SRIR encoded in spherical harmonics (SHs), we extract the DRIRs through plane wave decomposition for a set of angles distributed around the sphere 
\begin{equation}
y(n, \phi, \theta) = \mathbf{y}^T(\phi, \theta)\,\mathbf{s}(n),
\label{eq:drir_beam}
\end{equation}

\noindent where $\mathbf{s}$ is the SRIR, and $\mathbf{y}$ is a vector of SH transforms for a pair of azimuth and elevation angles $(\phi, \theta)$. Directional decay curves are calculated from a DRIR with \cite{alary_assessing_2019, berzborn_directional_2019, Alary_perceptual_2021}
\begin{equation}
\EDC(n, \omega, \phi, \theta) = 10\,\log_{10}\Bigg(\sum_{i=n}^{N} \big(y(i, \omega, \phi, \theta)\big)^2\Bigg),
\label{eq:DEDC2}
\end{equation}

\noindent where $\omega$ is the center frequency of a band in a filter bank.

To analyze the directional decay characteristics contained in the DRIR, we first calculate a mean energy decay curve from individual directional decay curves through
\begin{equation}
\overline{\EDC}(n, \omega) = \frac{1}{K} \sum_{k=1}^{K} \EDC(n, \omega, \phi_k, \theta_k),
\end{equation}

\noindent which is then subtracted from specific directions to obtain a measure of the energy decay deviation (EDD) \cite{alary_assessing_2019, Alary_perceptual_2021}, defined as
\begin{equation}
\EDD(n, \phi, \theta) = \EDC(n, \phi, \theta) - \overline{\EDC}(n).
\end{equation}

\begin{figure}[!t]
\centerline
{\includegraphics[trim=0cm 0cm 0cm 0cm, width=1.0\columnwidth]{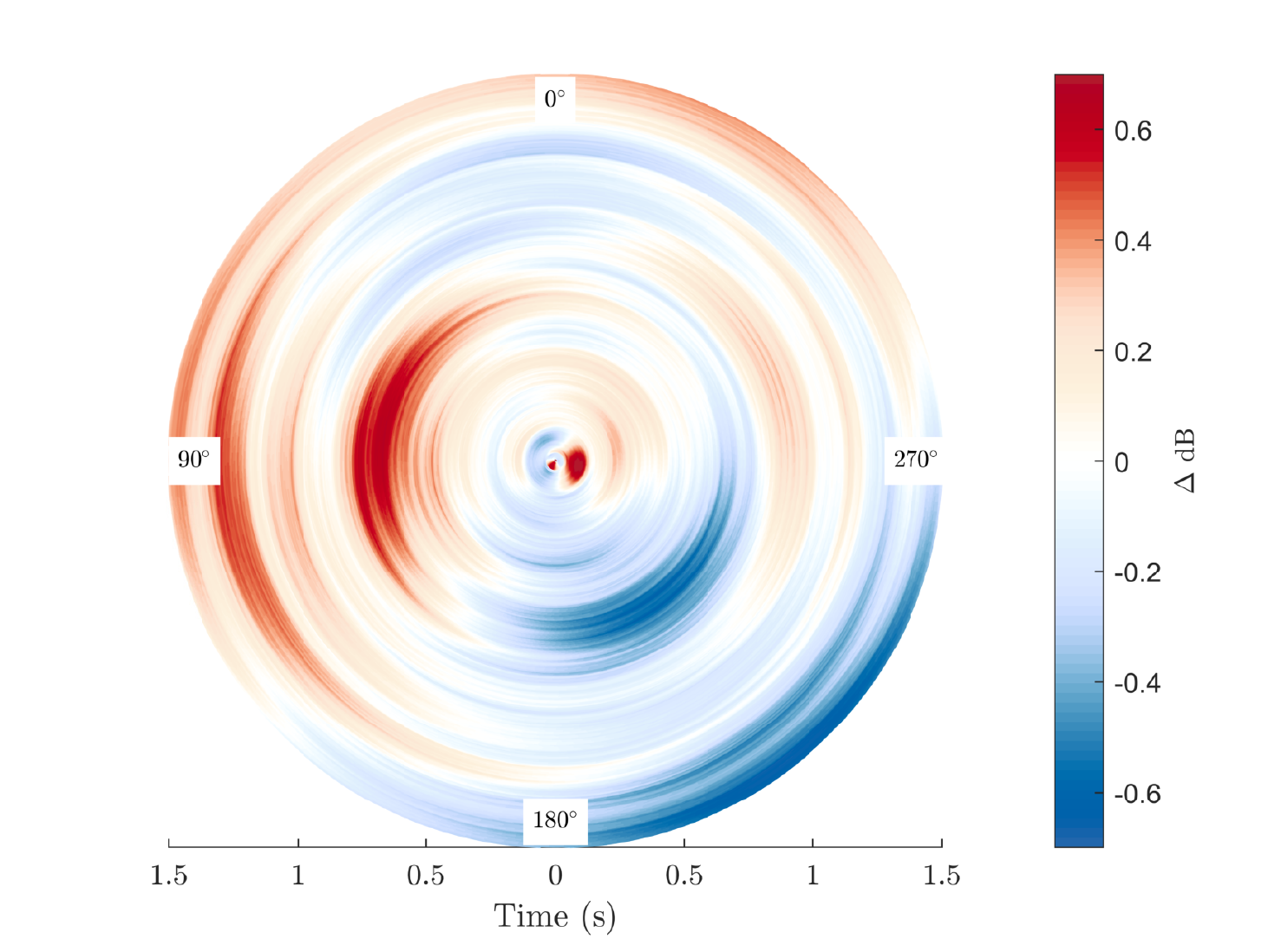}}
\caption{\it EDD analysis of a measured SRIR, illustrating directional decay deviations on the lateral plane. The radius is the time axis and the areas in red represent directions with more energy in the decay, while directions in blue have less.}
\label{fig:EDD}
\end{figure}

In Fig.~\ref{fig:EDD}, the EDD analysis is performed on the lateral plane of a captured SRIR. The SRIR is taken from the data set detailed in the following section. The red color is used to highlight directions with more energy in the decay when compared to the mean, while blue represents directions with less energy.

\subsection{Directional Feedback Delay Network}

\begin{figure*}
\centering



\begin{tikzpicture}[scale=0.9, every node/.style={scale=0.90}]
	\matrix[row sep=3.0mm, column sep=4.2mm]
	{
	

		& & & & & & 
		\node[dspsquare, minimum size=8em] 		(a02) {$\mathbfcal{A}$}; 	&
		& & & &	& \\

		
		\node[dspnodeopen,dsp/label=left] 		(m01) {$\mathbf{x}(n)$};    	&
		\node[dspnodefull]                 		(m02) {};          	& 
		\node[dsphadamard, dsp/label=above]  		(m03) {$\mathbf{b_1}$}; & & &
		\node[dspadder]  						(m04) {};    		&
		\node[dspsquare, minimum width=5em] 	(m05) {$\mathbf{z^{-m_1}}$}; &
		\node[dspsquare, minimum width=3.1em]     (m06) {$\mathbf{G}_\mathbf{1}(z)$};          &   	& & & 
		\node[dsphadamard, dsp/label=above]  		(m07) {$\mathbf{c_1}$};    	& 
		\node[dspadder]  						(m08) {};    		& &
		\node[dspnodeopen,dsp/label=right] (m09) {$\mathbf{y}(n)$};    		\\
		 
		\node[coordinate]                  		(m11) {};          	& 
		\node[dspnodefull]                 		(m12) {};          	& 
		\node[dsphadamard, dsp/label=above]  		(m13) {$\mathbf{b_2}$};    	&
		&
		\node[dspadder]  						(m14) {};    		&
		&
		\node[dspsquare, minimum width=8em] 	(m15) {$\mathbf{z^{-m_2}}$}; &
		& 
		\node[dspsquare, minimum width=3.1em]     (m16) {$\mathbf{G}_\mathbf{2}(z)$};  & & &
		\node[dsphadamard, dsp/label=above]  		(m17) {$\mathbf{c_2}$};    	&
		\node[dspadder]  						(m18) {};    		&
		\\
		& 
		\node[label](mInB){\vdots}; & 
		\node[label](){\vdots}; & & & & 
		\node[label](){\vdots}; & & &
		\node[label](){$\ddots$}; & &
		\node[label](){\vdots}; &
		\node[label](mOutB){\vdots}; \\
		
		\node[coordinate]                  		(m21) {};          	&
		\node[coordinate]                 		(m22) {};          	& 
		\node[dsphadamard, dsp/label=above]  		(m23) {$\mathbf{b}_N$};    	&
		\node[dspadder]  						(m24) {};    		&
		& &
		\node[dspsquare, minimum width=7em] 	(m25) {$\mathbf{z}^{-m_N}$}; &
		& & &
		\node[dspsquare, minimum width=3.1em]     (m26) {$\mathbf{G}_N(z)$}; &   
		\node[dsphadamard, dsp/label=above]  		(m27) {$\mathbf{c}_N$};    	&
		\node[dspadder]  						(m28) {};    		&
		\\

		
		
		\node[coordinate]                  		(m31) {};          	&
		\node[coordinate]                 		(m32) {};          	&
        \node[dsphadamard, dsp/label=above] 	    (m33) {$\mathbf{d}$};        &
		\node[coordinate]  						(m34) {};    		&
		& &
		\node[coordinate]  						(m35) {};    		&
		& &
		\node[coordinate]     					(m36) {};           &
		\node[coordinate]  						(m37) {};    	    &&
		\node[coordinate]  						(m38) {};    		&
		\\
	};
	
	\foreach \r in {0, ...,2}
	{
	    

		\foreach \i [evaluate = \i as \j using int(\i+1)] in {2,...,7}
		{
		\begin{scope}[start chain]
			\chainin (m\r\i);
			\ifthenelse{\i = 7 \AND \r = 2}
			{\chainin (m\r\j) [join=by dspconnthick];}
			{\chainin (m\r\j) [join=by dspconnthick];}
		\end{scope}
		}
	}

	\foreach \r [evaluate = \r as \rr using int(\r+1)] in {0, ..., 2}
	{
		\draw[dspconnthick] (m\r6) |-  ($(a02.north east)!\rr/5!(a02.south east)$);
		\draw[dspconnthick] ($(a02.north west)!\rr/5!(a02.south west)$) -| (m\r4);
	}
	
	\draw[dspconnthick] (mOutB) -- (m18);
	\draw[dsplinethick] (mOutB) -- (m28);
	\draw[dsplinethick] (m12) -- (mInB);
	\draw[dsplinethick] (mInB) -- (m22);
	
	\draw[dspconnthick] (m18) -- (m08);
	\draw[dspconnthick] (m38) -- (m28);
	
	\draw[dspconnthick] (m32) -- (m33);
	
	\draw[dsplinethick] (m12) -- (m02);
	\draw[dsplinethick] (m32) -- (m22);
	
	\draw[dsplinethick] (m01) -- (m02);
	
	\draw[dsplinethick] (m33) -- (m38);
	
    
    \begin{scope}[decoration={
         markings, mark= at position 0.6em with {\node {/};\node[below=1pt]{$K$};}}]
        \draw [dsplinethick,line width=\dsplinewidththick,postaction=decorate] (m01.east|-m02) -- node[below=1pt] {} (m02);
    \end{scope}
    
    \begin{scope}[decoration={
         markings, mark= at position 0.6em with {\node {/};\node[below=1pt]{$K$};}}]
        \draw [dspconnthick,line width=\dsplinewidththick,postaction=decorate] (m08.east|-m09) -- node[below=1pt] {} (m09);
    \end{scope}
	
\end{tikzpicture}

\caption{Flow diagram of the DFDN \cite{alary_frequency-dependent_2020}. The delay paths are organized into delay groups. The thick lines represent multichannel signal paths.}
\label{fig:dfdnblockdiagram}
\end{figure*}
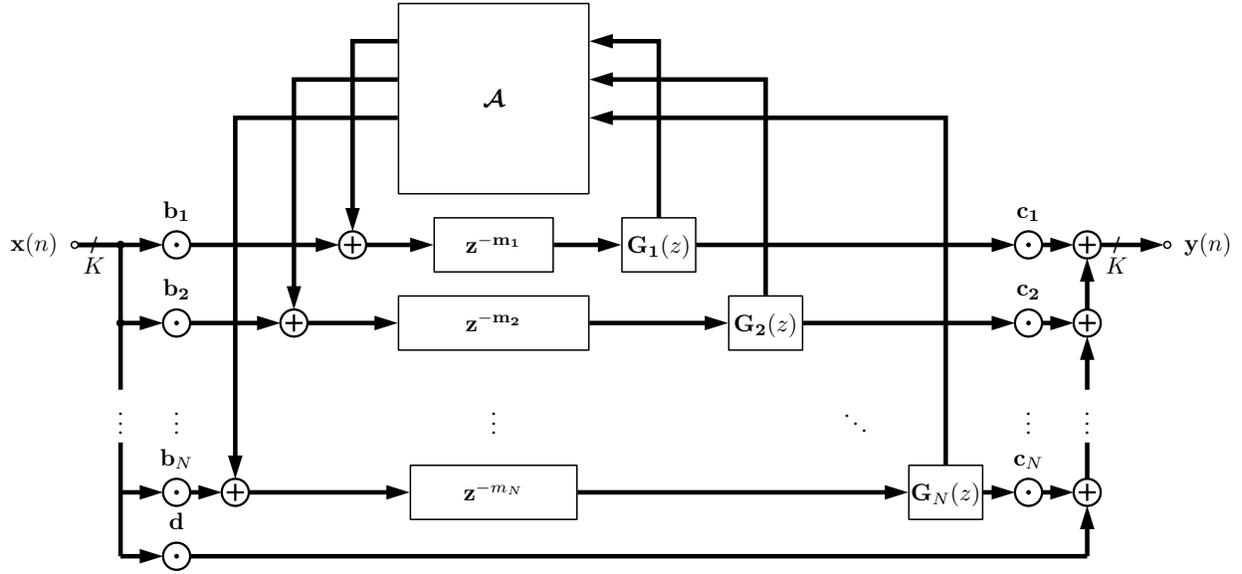

A DFDN \cite{alary_directional_2019, alary_frequency-dependent_2020} is a delay network reverberator which extends the design of an FDN \cite{jot_digital_1991} to produce multichannel reverberation with directional decay characteristics. In a DFDN, each delay path contains a group of $K$ delay lines to carry a multichannel signal (Fig.~\ref{fig:dfdnblockdiagram}). In this method, the channels may be defined as SH \cite{alary_directional_2019} or incident plane-waves distributed around a sphere \cite{alary_frequency-dependent_2020}. In this article, we use the plane-wave representation. The transfer function of the reverberator is defined as \cite{alary_frequency-dependent_2020}

\begin{equation}
	\mathbf{y}(z) = \sum_{i=1}^{N} \mathbf{c}_i^\mathrm{T}  \mathbf{G}_i(z)\, \mathbf{s}_{i}(z) + \mathbf{d} \odot \mathbf{x}(z),
\end{equation}

\begin{equation}
    \mathbf{s}_{i}(z) = \mathbf{Z}_i(z)\Bigg[\mathbf{b}_i \odot \mathbf{x}(z) + \sum_{j=1}^{N} A_{ij}\mathbf{G}_j(z) \mathbf{s}_{j}(z)\Bigg],
    \label{eq:dfdn2}
\end{equation}

\noindent where $\odot$ denotes element-wise multiplication (Hadamard product) and vectors $\mathbf{s}(z)$, $\mathbf{b}$, and $\mathbf{c}$ are
\begin{equation}
\mathbf{s}_i(z) = 
\begin{pmatrix}
s_{i,1}(z) \\
s_{i,2}(z) \\
\vdots \\
s_{i,K}(z)
\end{pmatrix},\,\mathbf{b}_i = 
\begin{pmatrix}
b_{i,1} \\
b_{i,2} \\
\vdots \\
b_{i,K}(z)
\end{pmatrix} ,\,\mathbf{c}_i = 
\begin{pmatrix}
c_{i,1} \\
c_{i,2} \\
\vdots \\
c_{i,K}
\end{pmatrix}, 
\end{equation}

\noindent representing the $K$ channels of the $i$th delay group. The matrices $\mathbf{Z}_i(z)$, and $\mathbf{G}_i(z)$ are defined as

\begin{equation}
\mathbf{Z}_i(z) = 
\begin{pmatrix}
z^{-m_{i,1}} & 0 & \cdots & 0 \\
0 & z^{-m_{i,2}} & \cdots & 0 \\
\vdots & \vdots & \ddots & \vdots \\
0 & 0 & \cdots & z^{-m_{i,K}} 
\end{pmatrix},
\end{equation}
\noindent and
\begin{equation}
\mathbf{G}_i(z) = 
\begin{pmatrix}
G_{i,1}(z)& 0 & \cdots & 0  \\
0 & G_{i,2}(z) & \cdots & 0  \\
\vdots & \vdots & \ddots & \vdots \\
0 & 0 & \cdots & G_{i,K}(z)
\end{pmatrix},
\end{equation}

\noindent where individual $G_{i,j}(z)$ contains the transfer function of an absorbent filter for the $j$th direction. 

To specify these filters, the reverberator uses a set of direction- and frequency-dependent decay times parameters $T_{60}(\omega, \phi, \theta)$. These decay times are used to calculate the necessary per-sample attenuation in the system using

\begin{equation}
g_\mathrm{dB}(\omega, \phi, \theta) = \frac{-60}{T_{60}(\omega, \phi, \theta)f_\mathrm{s}}, 
\label{eq:dbattenuation}
\end{equation}

\begin{equation}
g_\mathrm{lin}(\omega, \phi, \theta) = 10^{\frac{g_\mathrm{dB}(\omega, \phi, \theta)}{20}},
\end{equation}

\noindent which are then raised to the power of individual delay lengths in the system, to calculate the required attenuation at the output of each delay lines. The gain vector for a group of delay lines is defined as

\begin{equation}
\DeltaPerSampleAtt(\omega, \phi, \theta) = (g_\mathrm{lin}(\omega, \phi, \theta))^{\mathbf{m}_{i}},
\label{eq:rec_weight2}
\end{equation}

\noindent where $\mathbf{m}_{i}$ is the delay length vector of the $i$th delay group.

These gain values are used to specify a graphic equalizer after each delay lines, which are formed by a cascade of $L-1$ second-order IIR peak filters and a high-shelf filter \cite{prawda_improved_2019}: 
\begin{equation}
G(z) = G_0\prod\limits_{l=1}^L G_l(z).
\label{eq:DFDN_EQ}
\end{equation} 
\noindent where $G_0$ is the overall broadband gain factor. 

\section{Proposed Method}

\begin{figure*}[!t]
\centerline
{\includegraphics[trim=0cm 0cm 0cm 0cm, width=1.8 \columnwidth]{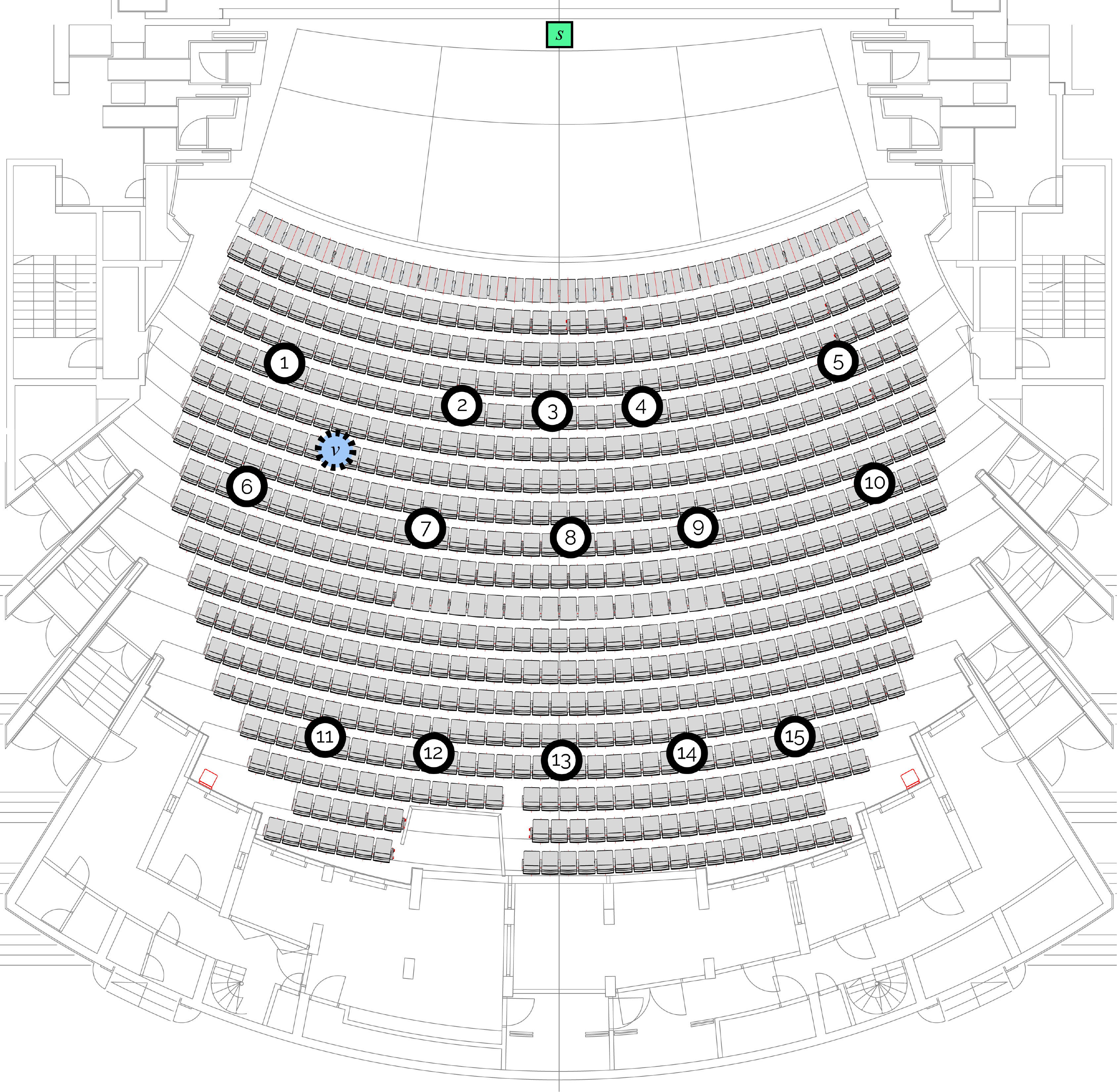}}
\caption{\it Main auditorium at the National Finnish Opera and Ballet in Helsinki, Finland. The positions of the loudspeaker ($S$) and microphones (circles numbered from 1 to 15) are shown.  The point $v$ is the location of a virtual point used to illustrate the interpolation of parameters in Fig.~\ref{fig:InterpolationTest}.}
\label{fig:Ooppera}
\end{figure*}

Due to the inhomogeneous and anisotropic characteristics of the sound field in an opera hall, which is induced by the coupled volumes it contains \cite{Massimo_opera_2016}, the proposed framework for reproducing their acoustics involves the capture of SRIRs using a spherical microphone array from multiple positions in the hall (Fig.~\ref{fig:Ooppera}). The resulting data set is later analyzed in order to extract a subset of key perceptual characteristics of the decaying sound field at each location. Using the analyzed information, we use a DFDN reverberator to auralize directional decay characteristics in 6DoF.
The proposed method adapts to any distribution of data points using interpolation, and the spatial resolution of the reverberator is chosen to satisfy computational constrains.

The decay parameters of the DFDN, which operates on a set of directions $(\phi, \theta)$ and bands with center frequency $\omega$, are specified by analyzing the measured SRIRs, as defined in Eq.~\eqref{eq:drir_beam} and Eq.~\eqref{eq:DEDC2}.

\subsection{Capturing the Data Set}

We captured a data set of SRIRs in the main auditorium of the Finnish National Opera and Ballet in Helsinki, Finland. For this experiment, we placed a 8430A Genelec loudspeaker in the center of the main stage of the auditorium, in front of the orchestral pit, which was opened. We acquired SRIRs from 15 microphone locations using a fourth order ambisonics microphone (Fig.~\ref{fig:Ooppera}). The SRIRs were measured using a sine sweep between 80\,Hz and 12\,kHz over 30\,s. The gain of the microphone was adjusted with sufficient headroom from the closest microphone position, and the microphone gain was kept constant throughout the recording.
The ventilation system was turned off to minimize the background noise. 

Depending of the desired application, it may be beneficial to vary the sound source location as well, to capture SRIRs containing information on the ERs from different locations. However, in this experiment, we were mainly interested in capturing different microphones positions.

In this section, we give an overview of the various parameters that are analyzed from the measured data set and how this information is used to specify the design of the delay network reverberator.

\subsection{Early Reflections}

Some of the more critical auditory cues of a room come from ERs \cite{Barron_early_1981}. One of several approaches may be used to extract or simulate the ERs for reproduction. In the proposed method, the ERs are processed separately in a parallel signal path, and the emphasis is on the late reverberation. As such, the method used for the reproduction of ERs may vary depending of the application. For instance, using individual SRIRs, it is possible to extract the direction of arrival (DOA) of individual reflections \cite{Tervo_doa_2015}, but interpolating between captured points may still be challenging. A more accurate representation of the reflective surfaces may be obtained by leveraging more data points, which is suitable for 6DoF sound reproduction \cite{Otto_doa_2021}. For real-time auralization, another method is to use geometric acoustics to simulate the sound propagation using the geometry of the room \cite{vorlander_auralization_2008, savioja_overview_2015}.

\subsection{Early Decay}

As the ERs are reproduced separately, the DFDN is designed to produce late reverberation. A mono input signal ($x$) is first copied to each delay group, but needs to be weighted. Indeed, an important characteristic of an IR is the early decay time (EDT), defined as the time required for the first 10\,dB of attenuation in an IR. However, in the case of an SRIR, the relative amplitude of different directions is also an important characteristic. For this reason, we chose to analyze the spatial distribution of energy from the beginning of the response by using the EDC value of the first sample ($\mathrm{EDC}_0$), corresponding to the total energy of a given DRIR. The analysis is performed on a 840-point grid distributed around the sphere \cite{Graf_tdesign_2010}.

\begin{figure}[!t]
\centerline
{\includegraphics[trim=0cm 0cm 0cm 0cm, width=1.0\columnwidth]{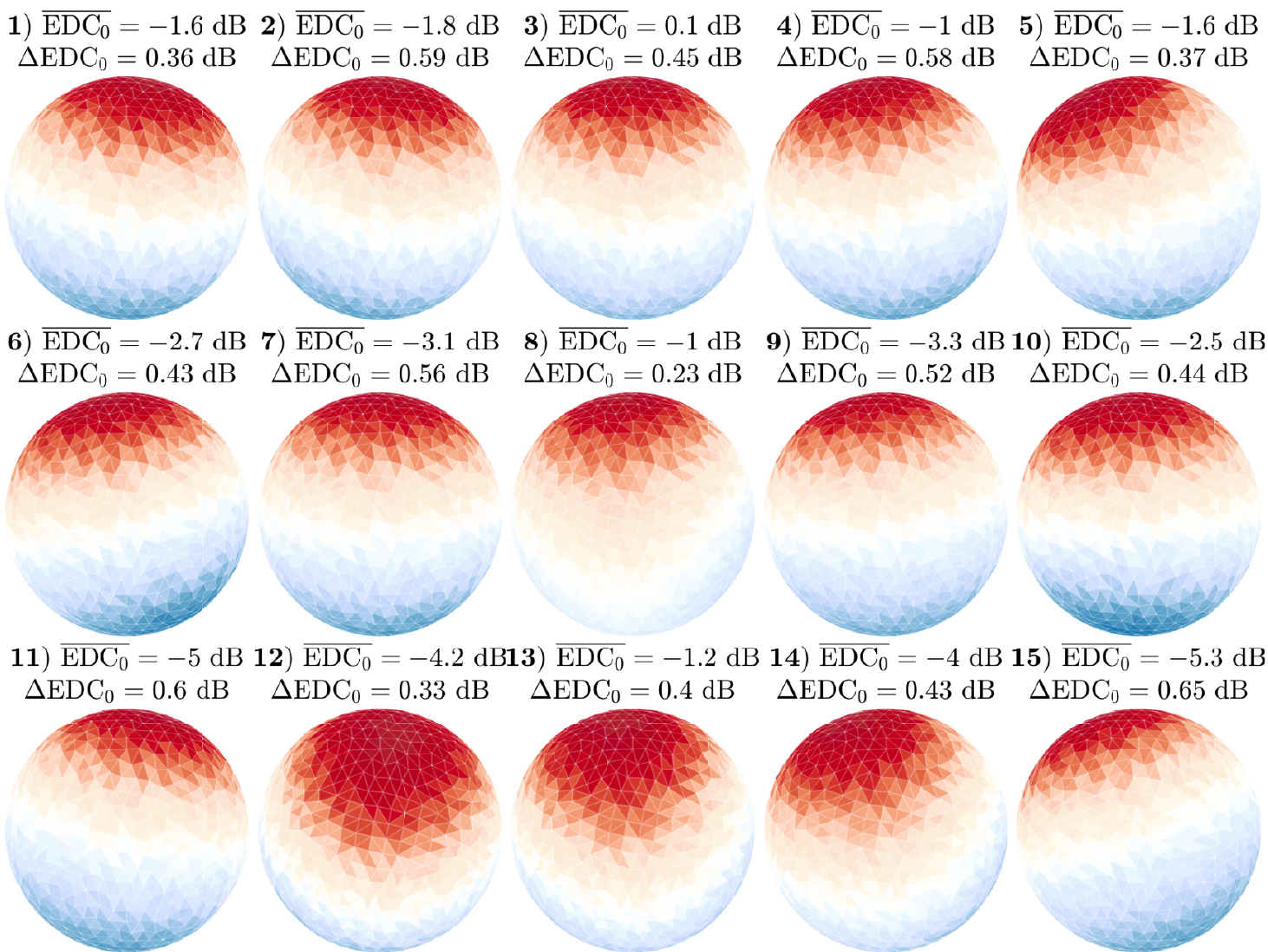}}
\caption{\it Distribution of $\EDC_0$ values on the top hemisphere, representing the total energy in the decay of the sound field from the various captured microphones locations. }
\label{fig:EarlyStuff}
\end{figure}

Figure \ref{fig:EarlyStuff} shows the directional distribution of values on the top hemisphere for each measurement, from perspective illustrated in Fig.~\ref{fig:Ooppera}. Unsurprisingly, due to the exponential decay, the energy predominantly comes from the direct sound. The $\EDC_0$ values are used to specify the input gain vector $\mathbf{b}_i$ of the DFDN, as defined in Eq.~\eqref{eq:dfdn2} and Fig.~\ref{fig:dfdnblockdiagram}.

\subsection{Directional Late Reverberation}

In order to specify the frequency- and direction-dependent absorbent filters in the recirculation path of the DFDN, we analyze the decay times $T_{60}(\omega, \phi, \theta)$, calculated from the $T_{30}$ of the DRIRs, at multiple bands $\omega$. In a FDN reverberator, it is common to use three frequency bands for low, mid, and high frequencies. The set of $K$ directions should be carefully selected as more directions will increase the computational complexity of the system, while a lower $K$ value will smooth out the directional features. In a previous study, a set of 12 directions equally distributed around the sphere has been shown to offer satisfactory results \cite{alary_frequency-dependent_2020}.

\begin{figure}[!t]
\centerline
{\includegraphics[trim=0cm 0cm 0cm 0cm, width=\columnwidth]{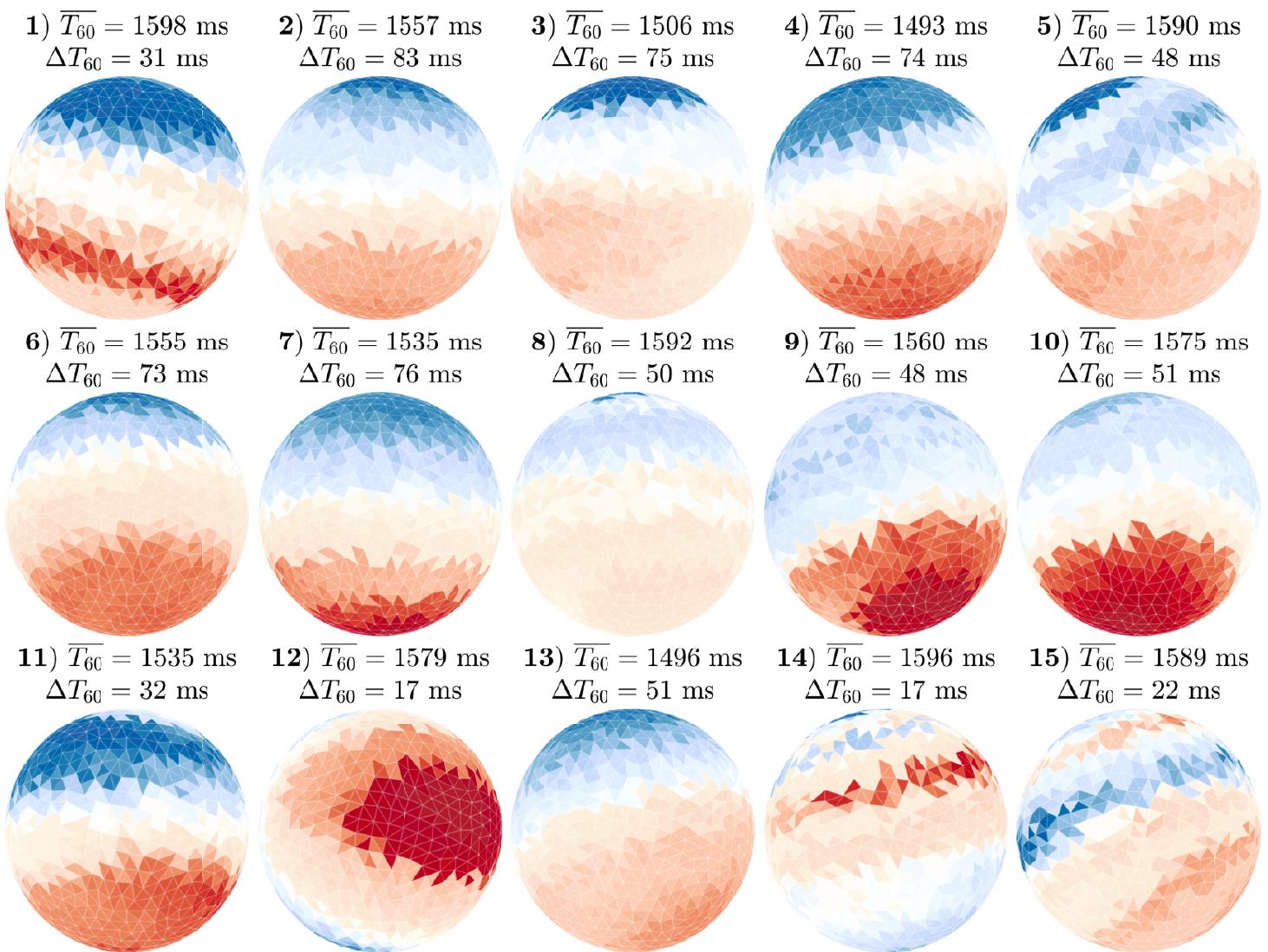}}
\caption{\it Estimated directional decay times for the captured SRIRs in the frequency range of $200\,$Hz-$800\,$Hz, the values are centered around a mean decay times with red representing longer decay times, and blue shorter ones. }
\label{fig:DT60_1}
\end{figure}

\begin{figure}[!t]
\centerline
{\includegraphics[trim=0cm 0cm 0cm 0cm, width=1.0\columnwidth]{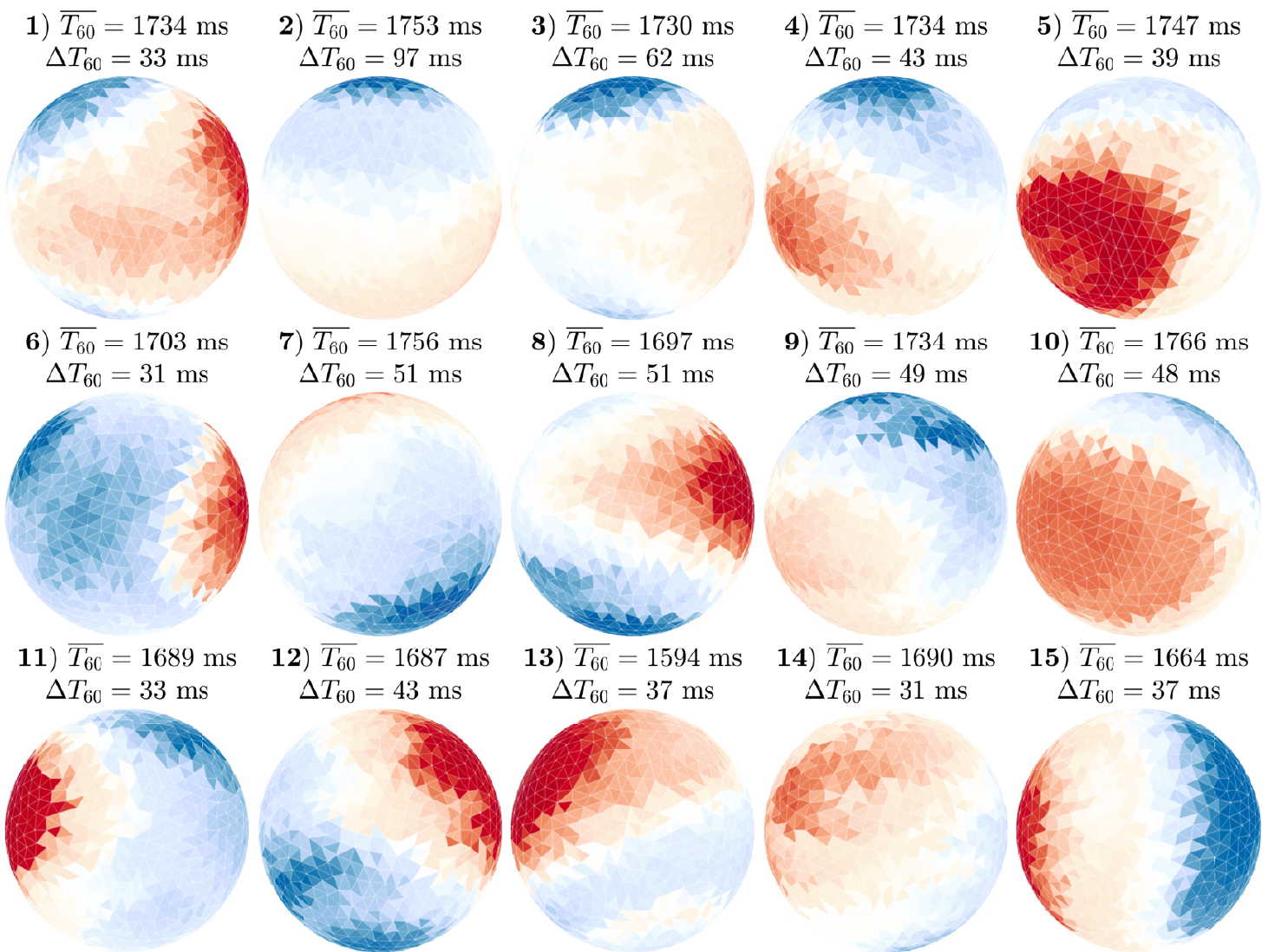}}
\caption{\it Estimated directional decay times for the captured SRIRs in the frequency range of $1\,$kHz-$2\,$kHz, the values are centered around a mean decay times with red representing longer decay times, and blue shorter ones.}
\label{fig:DT60_2}
\end{figure}

In Fig.~\ref{fig:DT60_1}, the spatial distribution of $T_{60}$ values are represented from the top hemisphere of each measured SRIRs, as represented in Fig.~\ref{fig:Ooppera}. Here, the results show the analysis for low frequencies, between 200\,Hz and 800\,Hz. For visualization, a mean $T_{60}$ ($\overline{T_{60}}$) is calculated for each captured SIR and white represents directions with a $T_{60}$ at this mean, while dark red and blue represent directions with respectively, longer or shorter $T_{60}$. The $\Delta T_{60}$ values corresponds to the difference between the mean and the maximum absolute value, meaning that dark red directions have a decay time of $T_{60} <= \overline{T_{60}}+\Delta T_{60}$ and darker blue directions have $T_{60} >= \overline{T_{60}}-\Delta T_{60}$. In this frequency range, the late reverberant energy is concentrated slightly more towards the back of the hall.

\begin{figure}[!t]
\centerline
{\includegraphics[trim=0cm 0cm 0cm 0cm, width=1.0\columnwidth]{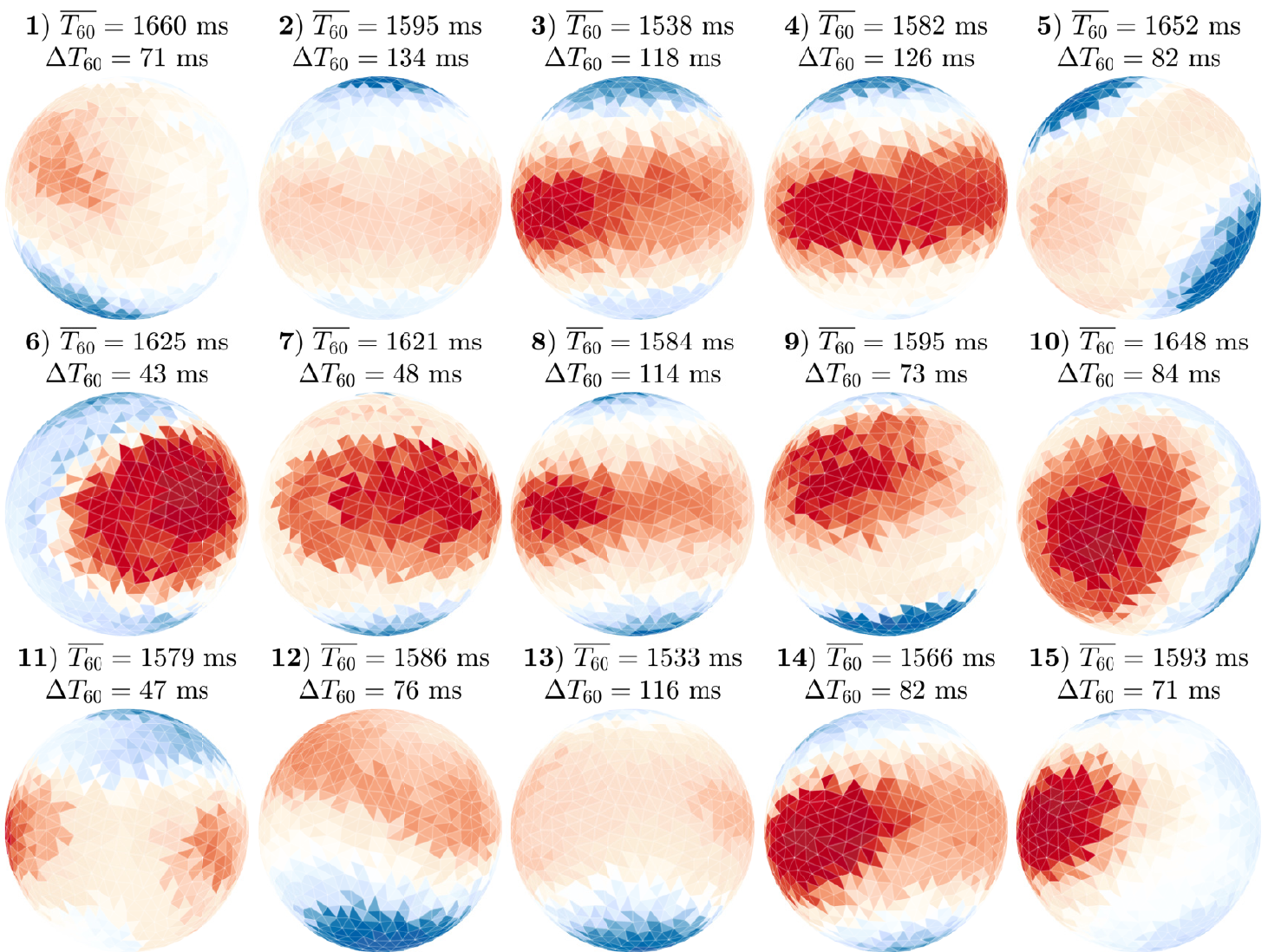}}
\caption{\it Estimated directional decay times for the captured SRIRs in the frequency range of $2\,$kHz-$6\,$kHz, the values are centered around a mean decay times with red representing longer decay times, and blue shorter ones. }
\label{fig:DT60_3}
\end{figure}

In Fig.~\ref{fig:DT60_2}, the same analysis was performed in the $1\,$kHz to $2\,$kHz band, which has a more varied distribution of late energy throughout the various microphone locations. The $\overline{T_{60}}$ is also slightly longer in this band compared to the other analyzed bands. Finally, Fig.~\ref{fig:DT60_3} shows that the late reverberation in the $2\,$kHz to $6\,$kHz frequency range is predominately incident from above the listening points.

Using the measured frequency- and direction-dependent decay times, we calculate the necessary attenuation and specify the three-band absorbent filters for each delay lines using Eq.~\eqref{eq:dbattenuation} through Eq.~\eqref{eq:DFDN_EQ}. The other parameters of the DFDN, including the orthogonal recirculating matrix and the length of individual delay line, may be perceptually tuned or randomized within an small range.

\subsection{Interpolation Between Measurement Points}

The analysis results of the various data points are used to specify some of the parameters of the DFDN when a listener is at these points. However, to reproduce any location in between these data points, an interpolation method is required to estimate the set of parameters. The interpolation occurs in the analyzed parameter space and not in the audio domain. As such, the main goal is to obtain interpolated values for $\mathrm{EDC}_0(\phi, \theta)$ and $T_{60}(\omega, \phi, \theta)$ in order to modulate the input gains $\mathbf{b}_i$ and absorbent filters $\mathbf{G}_i(z)$ in the DFDN. For a real-time implementation, the change of these values should occur over several audio frames to avoid any discontinuity and special consideration should be taken when modulating the IIR filters.

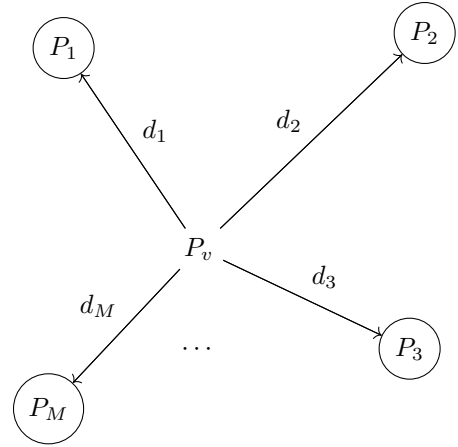
\begin{figure}[!t]
\centering
\begin{tikzpicture}
    \node[shape=circle,draw=black] (A) at (0,0) {$P_M$};
    \node[shape=circle,draw=black] (B) at (0.2, 4.8) {$P_1$};
    \node[shape=circle,draw=black] (C) at (5, 5) {$P_2$};
    \node[shape=circle,draw=black] (D) at (4.8, 0.8) {$P_3$};
    \draw (2, 2.4) node[anchor=north](S){$P_v$};
    \draw (2, 1) node[anchor=north](R){$\cdots$};
    \draw [double, ->] (S) edge (A) (S) edge (B) (S) edge (C) (S) edge (D);
    \draw (S) -- ++(A) node[midway, anchor=south east] {$d_M$};
    \draw (S) -- ++(B) node[midway, anchor=south west] {$d_1$};
    \draw (S) -- ++(C) node[midway, anchor=south east] {$d_2$};
    \draw (S) -- ++(D) node[midway, anchor=south west] {$d_3$};
\end{tikzpicture}
\caption{IDW interpolation method is based on the distances $d_1$...$d_M$ between a virtual point $P_v$ and its neighbouring data points $P_1$...$P_M$.}
\label{fig:IWD}
\end{figure}

Since our data set is relatively sparse, we use a simple interpolation technique for non-uniformly distributed points called the inverse distance weighting (IDW), which consists of weighting the values of surrounding points based on their distance to a given point \cite{Shepard_interpolation_1968}. 
We chose the IDW method since it is especially robust when using a data set distributed on a sparse and non-uniform grid, allowing a lot of flexibility with the measured data set. The IDW method is defined as
\begin{equation}
	w_i = \frac{\frac{1}{d_i}}{\sum_{i=1}^{M} \frac{1}{d_i}},
\end{equation}

\noindent where $d_i$ is the distance from a desired location to the $i$th of its $M$ nearest surrounding data points. Using these weights, the decay times of individual directions are obtained using a weighted average of the surrounding points through
\begin{equation}
	\mathbf{T_{60}} = \sum_{i=1}^{M} w_i \mathbf{T_{60}^{i}},
\end{equation}

\noindent where $\mathbf{T_{60}^{i}}$ is a vector containing the directional decay times for a given point.

\begin{figure}[!t]
\centerline
{\includegraphics[trim=0cm 0cm 0cm 0cm, width=0.9\columnwidth]{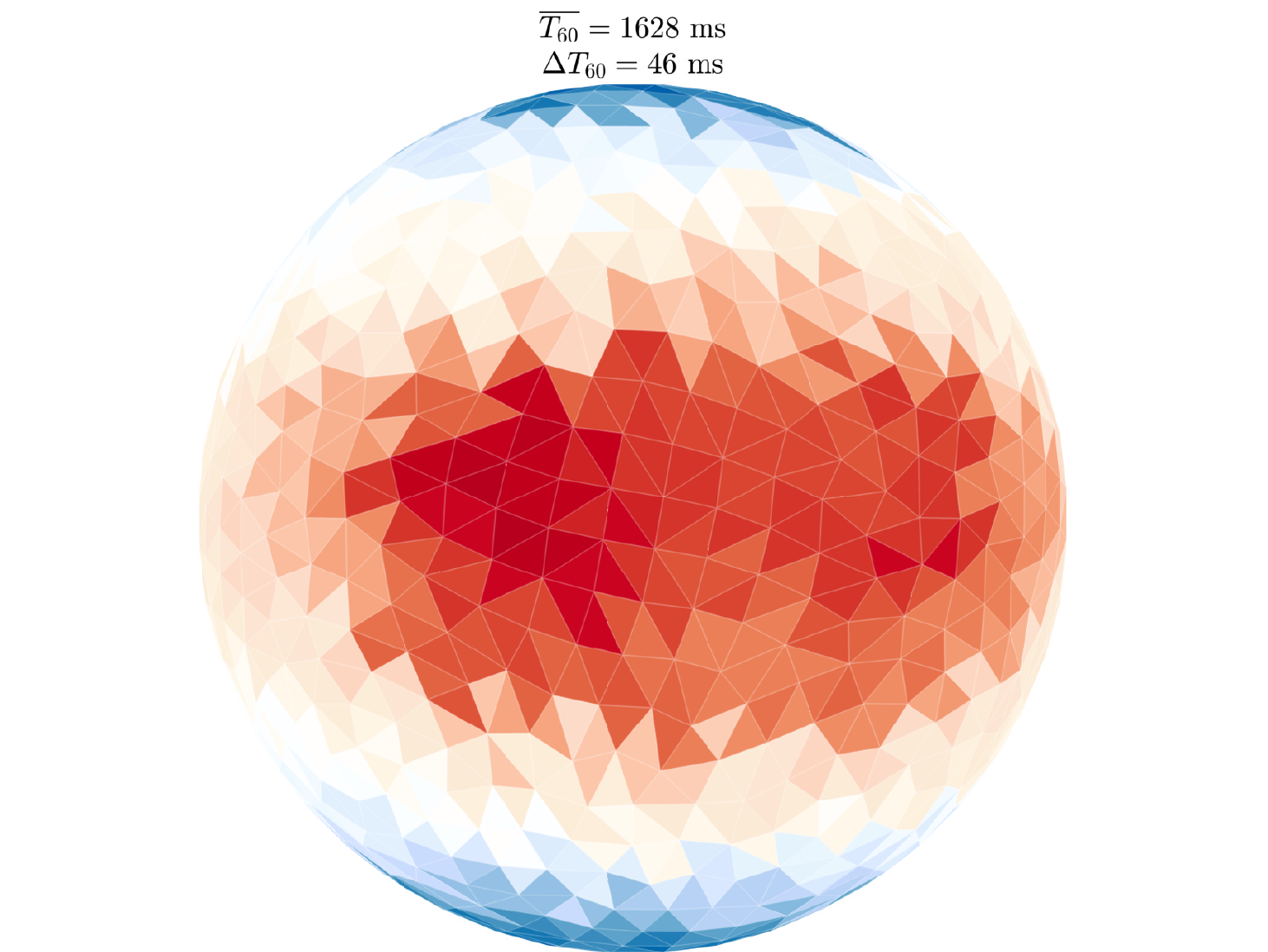}}
\caption{\it Interpolation results of directional reverberation times in the $2\,$kHz-$6\,$kHz frequency range for a virtual point located between data points 1, 2, 6, and 7, cf. Fig.~\ref{fig:DT60_3}. }
\label{fig:InterpolationTest}
\end{figure}

In Fig.~\ref{fig:InterpolationTest}, the interpolated decay times are given for the virtual point labeled $\varv$ in Fig.~\ref{fig:Ooppera}. The interpolation was performed on the data analysed in the $2\,$kHz-$6\,$kHz frequency range and the distances between $\varv$ and its four surrounding data points (1, 2, 6, and 7) were measured in meters as 2.909, 3.926, 2.883, and 3.608.

Since this performs a weighted averaged of the values, this interpolation approach may hide peaks and valleys in between the analyzed data points. As such, the main purpose of this interpolation is to modulate the gains in the reverberator and provide a smooth transition between the measured data points for auralization but it does not represent an accurate measure of the sound field at that location. A denser set of measurement points is likely to improve the accuracy. 

\subsection{Output Analysis of a DFDN}

\begin{figure}[!t]
\centerline
{\includegraphics[trim=0cm 0cm 0cm 0cm, width=1.0\columnwidth]{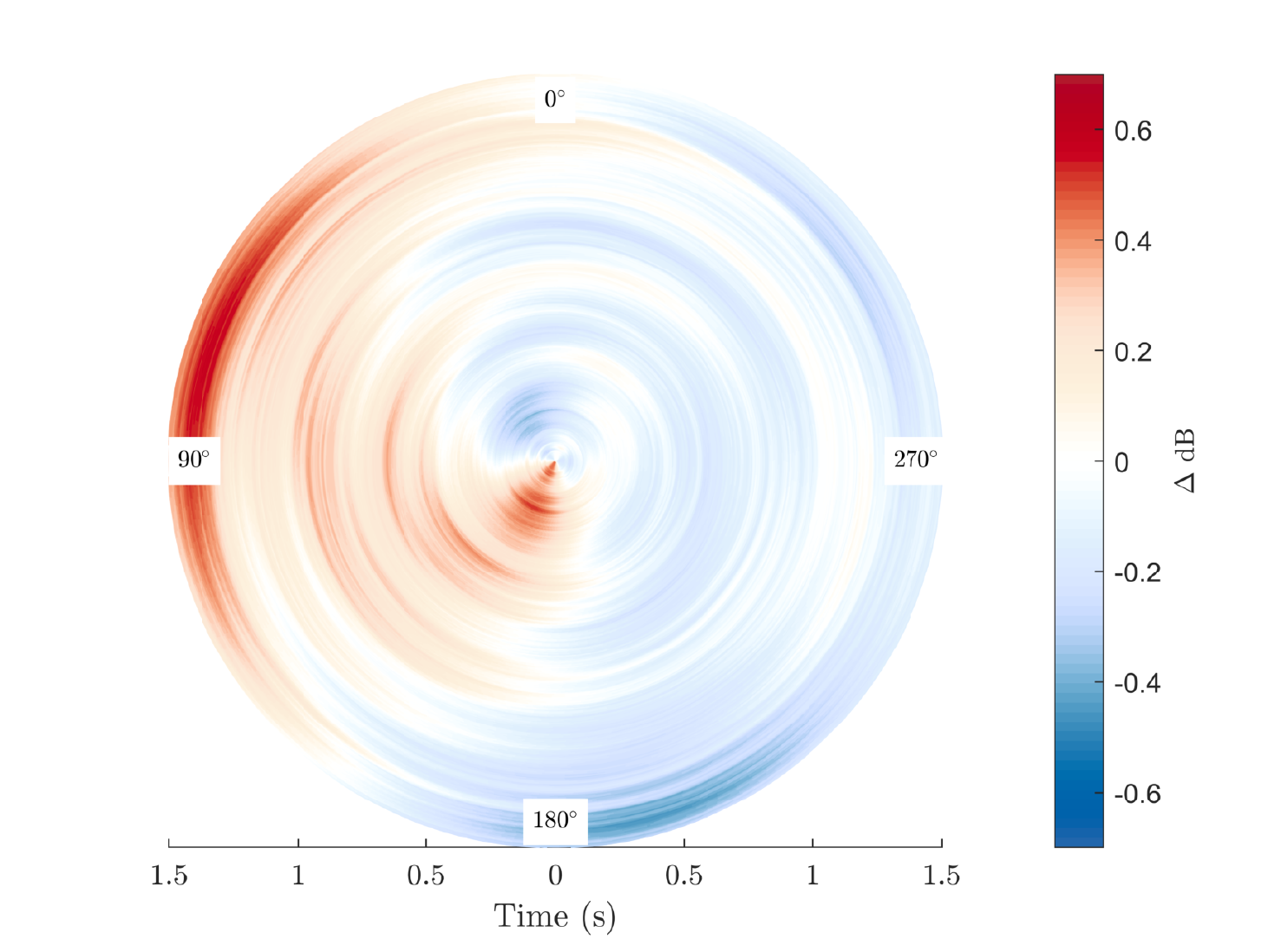}}
\caption{\it EDD analysis of a simulated SRIR using a DFDN reverberator, illustrating directional decay deviations on the lateral plane (cf.~Fig.~\ref{fig:EDD}). The radius is the time axis and the areas in red represent directions with more energy in the decay, while directions in blue have less.}
\label{fig:EDD_DFDN}
\end{figure}

Finally, in Fig.~\ref{fig:EDD_DFDN}, we analyze the EDD of an artificial SRIR generated using a DFDN, encoded in third-order Ambisonics from $K=24$ directions and 16 delay groups. The individual delay lengths and the recirculating matrix were randomized. The overall characteristics of the decay are reproduced while some smoothing occurs due in part to the ambisonics processing. The output of this reverberator may be encoded for binaural sound reproduction \cite{alary_frequency-dependent_2020}.

\section{Conclusion}
In conclusion, we presented a framework to capture, analyze, and reproduce the acoustics of a hall in 6DoF using a data set of measured SRIRs. The ERs are reproduced separately using pre-existing methods, while the late reverberation is reproduced using a DFDN reverberator. Using a data set of SRIRs, key perceptual parameters are extracted to specify the design of the artificial reverberator, capable of reproducing frequency- and direction-dependent decay characteristics. 

The values of these parameters are interpolated to provide a smooth transition between data points during sound reproduction. The interpolation method used offers flexibility on the measurement grid used for the captured data set, and the reverberation algorithm supports lower spatial resolution to satisfy computational constrains. Using the proposed framework, the directional reverberation of historical halls, such as the auditorium at the National Finnish Opera and Ballet in Helsinki, may be analyzed in more details while the DFDN provides an efficient means for their reproduction. 

Future work includes exploring different interpolation methods and extending the data set to a denser measurement grid in order to assess the interpolation accuracy. More historic halls will also be captured to study the directional characteristics present in real spaces. 

\section*{Acknowledgment}

The authors would like to thank Michael McCrea, Juha-Matti Vuo, and Timo Tuovila for their help and invaluable support in capturing impulse responses at the Finnish National Opera and Ballet.


\bibliographystyle{ieeetr}
\bibliography{DFDN}

\begin{thebibliography}{10}

\bibitem{Barron_ER_1971}
M.~Barron, ``The subjective effects of first reflections in concert
  halls---{The} need for lateral reflections,'' {\em Journal of Sound and
  Vibration}, vol.~15, pp.~475--494, Apr. 1971.

\bibitem{Schroeder_simulation_1970}
M.~R. Schroeder, ``Digital simulation of sound transmission in reverberant
  spaces,'' {\em J. Acoust. Soc. Am.}, vol.~47, pp.~424--431, Feb. 1970.

\bibitem{polack_playing_1993}
J.-D. Polack, ``Playing billiards in the concert hall: {The} mathematical
  foundations of geometrical room acoustics,'' {\em Appl. Acoust.}, vol.~8,
  pp.~235 -- 244, Feb. 1993.

\bibitem{nolan_isotropy_2020}
M.~Nolan, M.~Berzborn, and E.~Fernandez-Grande, ``Isotropy in decaying
  reverberant sound fields,'' {\em J. Acoust. Soc. Am.}, vol.~148,
  pp.~1077--1088, Aug. 2020.

\bibitem{Massimo_opera_2016}
M.~Garai, S.~De~Cesaris, F.~Morandi, and D.~D’Orazio, ``Sound energy
  distribution in {Italian} opera houses,'' {\em Proceedings of Meetings on
  Acoustics}, vol.~28, no.~1, p.~015019, 2016.

\bibitem{gerzon_recording_1975}
M.~A. Gerzon, ``Recording concert hall acoustics for posterity,'' {\em J. Audio
  Eng. Soc.}, vol.~23, pp.~569--571, Sep. 1975.

\bibitem{farina_recording_2003}
A.~Farina and R.~Ayalon, ``Recording concert hall acoustics for posterity,'' in
  {\em Proceedings of the Audio Engineering Society 24th International
  Conference: Multichannel Audio, The New Reality}, Jun 2003.

\bibitem{Zotter_Aur_2020}
K.~M\"{u}ller and F.~Zotter, ``Auralization based on multi-perspective
  ambisonic room impulse responses,'' {\em Acta Acustica}, vol.~4, p.~18, 11
  2020.

\bibitem{Noisternig_framework_2008}
M.~Noisternig, B.~F.~G. Katz, S.~Siltanen, and L.~Savioja, ``Framework for
  real-time auralization in architectural acoustics,'' {\em Acta Acustica
  united with Acustica}, vol.~94, no.~6, pp.~1000--1015, 2008.

\bibitem{Katz_Reconstruction_2020}
B.~F.~G. Katz, D.~Murphy, and A.~Farina, ``Exploring cultural heritage through
  acoustic digital reconstructions,'' {\em Physics Today}, vol.~73, no.~12,
  pp.~32--37, 2020.

\bibitem{jot_digital_1991}
J.-M. Jot and A.~Chaigne, ``Digital delay networks for designing artificial
  reverberators,'' in {\em Proc. Audio Eng. Soc. 90th Conv.}, (Paris, France),
  Feb. 1991.

\bibitem{valimaki_fifty_2012}
V.~V\"{a}lim\"{a}ki, J.~D. Parker, L.~Savioja, J.~O. Smith, and J.~S. Abel,
  ``Fifty years of artificial reverberation,'' {\em IEEE Trans. Audio, Speech,
  Lang. Process.}, vol.~20, pp.~1421--1448, Jul. 2012.

\bibitem{Vermeulen_stereo_1958}
R.~Vermeulen, ``Stereo-reverberation,'' {\em J. Audio Eng. Soc.}, vol.~6,
  pp.~124--130, Apr. 1958.

\bibitem{schroeder_colorless_1961}
M.~R. Schroeder and B.~F. Logan, ``'{Colorless}' artificial reverberation,''
  {\em J. Audio Eng. Soc.}, vol.~9, pp.~192--197, Jul. 1961.

\bibitem{Waterhouse_Interference_1955}
R.~V. Waterhouse, ``Interference patterns in reverberant sound fields,'' {\em
  J. Acoust. Soc. Am.}, vol.~27, pp.~247--258, Mar. 1955.

\bibitem{alary_assessing_2019}
B.~Alary, P.~Mass\'{e}, V.~V\"{a}lim\"{a}ki, and M.~Noisternig, ``Assessing the
  anisotropic features of spatial impulse responses,'' in {\em Proc. {EAA}
  {Spatial} {Audio} {Signal} {Processing} {Symposium}}, (Paris, France),
  pp.~43--48, Sep. 2019.

\bibitem{nolan_wavenumber_2018}
M.~Nolan, E.~Fernandez-Grande, J.~Brunskog, and C.-H. Jeong, ``A wavenumber
  approach to quantifying the isotropy of the sound field in reverberant
  spaces,'' {\em J. Acoust. Soc. Am.}, vol.~143, pp.~2514--2526, Apr. 2018.

\bibitem{berzborn_directional_2019}
M.~Berzborn, M.~Nolan, E.~Fernandez-Grande, and M.~Vorl\"{a}nder, ``On the
  directional properties of energy decay curves,'' in {\em Proc. 23rd {Int}.
  {Cong}. of {Acoustics}}, (Aachen, Germany), Sep. 2019.

\bibitem{Berzborn_dsfda_2021}
M.~Berzborn and M.~Vorl\"{a}nder, ``Directional sound field decay analysis in
  performance spaces,'' {\em Building Acoustics}, vol.~0, Jan. 2021.

\bibitem{romblom_perceptual_2016}
D.~Romblom, C.~Guastavino, and P.~Depalle, ``Perceptual thresholds for
  non-ideal diffuse field reverberation,'' {\em J. Acoust. Soc. Am.}, vol.~140,
  pp.~3908--3916, Nov. 2016.

\bibitem{Alary_perceptual_2021}
B.~Alary, P.~Mass\'{e}, S.~J. Schlecht, M.~Noisternig, and V.~V\"{a}lim\"{a}ki,
  ``Perceptual analysis of directional late reverberation,'' {\em The Journal
  of the Acoustical Society of America}, vol.~149, pp.~3189--3199, May 2021.

\bibitem{alary_directional_2019}
B.~Alary, A.~Politis, S.~J. Schlecht, and V.~V\"{a}lim\"{a}ki, ``Directional
  feedback delay network,'' {\em J. Audio Eng. Soc.}, vol.~67, pp.~752--762,
  Oct. 2019.

\bibitem{alary_frequency-dependent_2020}
B.~Alary and A.~Politis, ``Frequency-dependent directional feedback delay
  network,'' in {\em Proc. IEEE ICASSP-2020}, pp.~176--180, May 2020.

\bibitem{schroeder_new_1965}
M.~R. Schroeder, ``New method of measuring reverberation time,'' {\em J.
  Acoust. Soc. Am.}, vol.~37, pp.~409--412, Mar. 1965.

\bibitem{jot_analysissynthesis_1992}
J.-M. Jot, ``An analysis/synthesis approach to real-time artificial
  reverberation,'' in {\em Proc. {IEEE} {ICASSP}-92}, vol.~2, (San Francisco,
  CA), pp.~221--224, Mar. 1992.

\bibitem{masse_denoising_2020}
P.~Mass\'{e}, T.~Carpentier, O.~Warusfel, and M.~Noisternig, ``Denoising
  directional room impulse responses with spatially anisotropic late
  reverberation tails,'' {\em Appl. Sci.}, vol.~10, p.~1033, Feb. 2020.

\bibitem{Hopkins_Insulation_2007}
C.~Hopkins, {\em Sound Insulation}.
\newblock Oxford, UK: Butterworth-Heinemann, Jan. 2007.

\bibitem{prawda_improved_2019}
K.~Prawda, V.~V\"{a}lim\"{a}ki, and S.~J. Schlecht, ``Improved reverberation
  time control for feedback delay networks,'' in {\em Proc. Int. Conf. Digital
  Audio Effects (DAFx-19)}, (Birmingham, UK), Sep. 2019.

\bibitem{Barron_early_1981}
M.~Barron and A.~H. Marshall, ``Spatial impression due to early lateral
  reflections in concert halls: The derivation of a physical measure,'' {\em
  Journal of Sound and Vibration}, vol.~77, pp.~211--232, July 1981.

\bibitem{Tervo_doa_2015}
S.~Tervo and A.~Politis, ``Direction of arrival estimation of reflections from
  room impulse responses using a spherical microphone array,'' {\em IEEE/ACM
  Trans. Audio Speech Lang. Process.}, vol.~23, pp.~1539--1551, June 2015.

\bibitem{Otto_doa_2021}
O.~Puomio, N.~Meyer-Kahlen, and T.~Lokki, ``Locating image sources from
  multiple spatial room impulse responses,'' {\em Applied Sciences}, vol.~11,
  Mar. 2021.

\bibitem{vorlander_auralization_2008}
M.~Vorl\"{a}nder, {\em Auralization}.
\newblock Springer, 2008.

\bibitem{savioja_overview_2015}
L.~Savioja and U.~P. Svensson, ``Overview of geometrical room acoustic modeling
  techniques,'' {\em J. Acoust. Soc. Am.}, vol.~138, pp.~708--730, Aug. 2015.

\bibitem{Graf_tdesign_2010}
M.~Gr\"{a}f and D.~Potts, ``On the computation of spherical designs by a new
  optimization approach based on fast spherical {Fourier} transforms,'' {\em
  Numerische Mathematik}, vol.~119, pp.~699--724, Jan. 2010.

\bibitem{Shepard_interpolation_1968}
D.~Shepard, ``A two-dimensional interpolation function for irregularly-spaced
  data,'' in {\em Proceedings of the 23rd ACM National Conference}, ACM'68,
  (New York, NY, USA), p.~517–524, 1968.

\end{thebibliography}

\end{document}